\documentclass[sigconf, nonacm]{acmart}
\usepackage{algorithm}
\usepackage{algpseudocode}
\usepackage{amsmath}
\usepackage{caption}
\usepackage{graphicx}
\usepackage{subcaption}
\usepackage{balance}
\usepackage{multirow, makecell}
\usepackage{enumitem}

\newcommand{\probP}{\text{I\kern-0.15em P}}
\algrenewcommand\algorithmicrequire{\textbf{Input:}}
\algrenewcommand\algorithmicensure{\textbf{Output:}}

\newcommand{\far}{\textsf{GrASP}}

\newcommand{\centered}[1]{\begin{tabular}{c} #1 \end{tabular}}

\newcommand{\new}[1]{\textcolor{black}{#1}}

\begin{document}
\title{{\far}: A Generalizable Address-based Semantic Prefetcher for Scalable Transactional and Analytical Workloads}

\author{Farzaneh Zirak}
\orcid{0009-0005-1801-2565}
\affiliation{%
  \institution{University of Melbourne}
  \city{Melbourne}
  \country{Australia}
}
\email{fzirak@student.unimelb.edu.au}

\author{Farhana Choudhury}
\orcid{0000-0001-6529-4220}
\affiliation{%
  \institution{University of Melbourne}
  \city{Melbourne}
  \country{Australia}
}
\email{farhana.choudhury@unimelb.edu.au}

\author{Renata Borovica-Gajic}
\orcid{0000-0003-3503-4123}
\affiliation{%
  \institution{University of Melbourne}
  \city{Melbourne}
  \country{Australia}
}
\email{renata.borovica@unimelb.edu.au}

\begin{abstract}
  \textit{Data prefetching}—loading data into the cache before it is requested— is essential for reducing I/O overhead and improving database performance. While traditional prefetchers focus on sequential patterns, recent learning-based approaches, especially those leveraging data semantics, achieve higher accuracy for complex access patterns. However, these methods often struggle with today’s dynamic, ever-growing datasets and require frequent, timely fine-tuning. 
\new{Privacy constraints may also restrict access to complete datasets, necessitating prefetchers that can learn effectively from samples.}

\new{To address these challenges, we present {\far}, a learning-based prefetcher designed for both analytical and transactional workloads. {\far} enhances prefetching accuracy and scalability by leveraging logical block address deltas and combining query representations with result encodings. It frames prefetching as a context-aware multi-label classification task, using multi-layer LSTMs to predict delta patterns from embedded context. This delta modeling approach enables {\far} to generalize predictions from small samples to larger, dynamic datasets without requiring extensive retraining.  Experiments on real-world datasets and industrial benchmarks demonstrate that {\far} generalizes to datasets {$250\times$} larger than the training data, 
    achieving up to 45\% higher hit ratios, 60\% lower I/O time, and 55\% lower end-to-end query execution latency than existing baselines. On average, {\far} attains a 91.4\% hit ratio, a 90.8\% I/O time reduction, and a 57.1\% execution latency reduction.}

\end{abstract}

\maketitle

\begingroup\small\noindent\raggedright\\
The source code is made available at \url{https://github.com/fzirak/GrASP}.
\endgroup

\section{Introduction}

    \emph{Data prefetching} is a fundamental technique employed by database management systems (DBMS) to improve performance by reducing I/O time. A prefetcher anticipates future accesses and proactively loads relevant data into cache. Prior work has explored various prefetching techniques, from rule-based mechanisms to deep learning models~\cite{chen2021revisiting, battle2016prefetching, selep, opdenacker2007readahead, ki2000stride}, proving effective in diverse workloads.

 \new{Traditional prefetchers rely on sequential or locality patterns, while recent learning-based models better capture complex access behaviors~\cite{chen2021revisiting, battle2016prefetching, tauheed2012scout, selep}. Notably, semantic-based learning prefetchers achieve higher accuracy in capturing intricate and non-trivial patterns by leveraging data characteristics~\cite{battle2016prefetching, selep, tauheed2012scout}. However, state-of-the-art (SOTA) learning-based prefetchers often struggle to scale with today’s rapidly growing datasets. They fail to generalize to evolving workloads or modified datasets without timely and costly fine-tuning, which can degrade system responsiveness.}

    \new{This limitation is critical in modern data systems where workloads continuously evolve and datasets grow rapidly, demanding scalable solutions~\cite{ScientificDataManagementGray, ResearcherGuide}. Timely access to relevant information is essential in many applications~\cite{timelyanal3, timleyanal2, timelyanal4}, requiring prefetchers that scale and adapt without extensive retraining overhead.}

\new{These challenges are amplified in \textit{data exploration} scenarios, where users seek timely insights from (newly ingested) data. While some exploration tasks involve static analytical workloads, many require rapid analysis of frequently updated datasets~\cite{Analyst}. For example, analysts may monitor recent stock transactions to detect fraud~\cite{stockAnalyst}, or track social media streams for emerging trends~\cite{socialAnalyst}. In such cases, prefetchers may lack sufficient time to preprocess new batches or adapt prediction models, limiting their effectiveness.}

\new{Another constraint arises when the prefetcher cannot access the full dataset during training, due to either constant updates or more often privacy restrictions, as in medical or enterprise environments~\cite{privacyPreserving}.  Data owners often limit access to complete datasets and query workloads for confidentiality reasons, reducing the effectiveness of deep learning models. This calls for prefetchers that can train on limited samples while effectively generalizing across a much larger, unseen data space.}

   Given these constraints, we aim to design a prefetcher that accurately anticipates data accesses while meeting two core goals:
\begin{enumerate}[label=\textbf{\roman*.}, leftmargin=3em, labelsep=0.4em]
\vspace{-0.5em} 
    
    {\item{\textbf{Generalizable to larger datasets.}} A prefetcher must remain effective even when trained on a much smaller subset of the full deployment dataset. Upon deployment, it should adapt its predictions without requiring extensive retraining.}

    {\item{\textbf{Compatible with both analytical and transactional workloads.}} Modifying transactional workloads often introduces new data blocks or alters existing ones, changing block semantics. These shifts challenge SOTA prefetchers that rely on precomputed data semantics~\cite{selep} or restrict prefetching to blocks observed during training~\cite{selep, chen2021revisiting}. Frequent fine-tuning is often impractical, necessitating a prefetcher that can consider the full data space and adapt to changes with minimal adjustment.}

\end{enumerate}
\new{Recent memory prefetchers~\cite{2023sgdp, 2020delta_lstm} model the LBA delta, which is the difference between successive logical block address (LBA) requests, to predict access patterns across dynamic data spaces. This LBA-based modeling has also been adopted in several traditional~\cite{smith1978lookahead, opdenacker2007readahead} and learning-based~\cite{chen2021revisiting} database prefetchers.}
    
   \new{In contrast, semantic prefetchers~\cite{battle2016prefetching, tauheed2012scout, selep} have shown that leveraging data semantics instead of LBA information better captures dependencies between accessed data and improves prediction accuracy. However, they are limited to analytical queries and fail to generalize under transactional updates.}

    To address these gaps, we introduce {\far}—a learning-based prefetcher that combines LBA-delta (delta in short) modeling with semantic context to improve both scalability and accuracy. {\far} formulates prefetching as a contextual multi-label classification task, predicting future data accesses by forecasting delta values using recent query semantics and LBA information. It employs a multi-layer long short-term memory (LSTM) model to learn delta patterns from embedded semantic and LBA-based contexts.

\new{\textit{Semantic-based context.} {\far} incorporates data semantics by dynamically preprocessing and encoding data blocks using feature extraction techniques.  
   To avoid over-reliance on static block encodings, {\far} defines query semantics as a combination of query result encodings—aggregated from the accessed block encodings—and a query statement representation. This representation includes features such as query type, accessed tables, join conditions, and filter predicates.}
    
\textit{LBA-based context.} We introduce a table-based LBA abstraction to mitigate the effects of database growth on LBA and delta values. Additionally, we define an \emph{order-agnostic} delta to represent the \emph{set} of deltas associated with each query, independent of access order.

\new{{\far} constructs its input context by combining semantic and LBA features with metadata such as the last accessed tables and the number of deltas per query. Given a sequence of such contexts, it predicts the most probable deltas for the next query and identifies the corresponding candidate LBAs to prefetch.} Prefetching tasks often suffer from class imbalance due to skewed data access patterns, increasing the risk of overfitting to frequent classes. {\far} mitigates this issue by employing a custom loss function and applying dropout regularization to improve generalization.
    
    In this paper, we make the following contributions:
    \begin{itemize} [leftmargin=2.5em, labelsep=0.4em]
    \vspace{-0.3em}
           \item We introduce {\far}, a hybrid prefetcher that integrates semantic-aware features with delta modeling, using a table-based LBA abstraction and  order-agnostic delta formulation.

       \item We formulate prefetching as a contextual classification problem, \new{leveraging a novel integration of} query semantics and LBA information \new{to improve accuracy and generalizability}.

        \item \new{{\far} generalizes effectively, transferring learned delta patterns to significantly larger datasets with minimal tuning.}

       \item \new{Extensive experiments on real-world exploratory analytical workloads and industrial transactional benchmarks show that {\far} achieves an average hit ratio of 91.4\%, along with a 90.8\% I/O time reduction and a 57.1\% reduction in execution latency. Compared to state-of-the-art prefetchers, {\far} improves hit ratio by up to 17\%, reduces I/O time by up to 36\%, and lowers execution latency by up to 28\% in analytical workloads; in transactional workloads, improvements reach up to 45\%, 60\%, and 55\%, respectively.}

    \end{itemize}

 \vspace{-0.9em}  
\section{Background and Related Work}\label{sec:background_motiv}
    We start this section with the preliminaries in \S \ref{sec:preliminaries} and the prefetching problem definition in \S \ref{sec:problem_formulation}, followed by an overview of existing prefetching systems in \S \ref{sec:existing_prefetcher} and query encoding methods in \S \ref{sec:query_encoding}.
    \vspace{-0.5em}  
    \subsection{Preliminaries}\label{sec:preliminaries}
        \new{The challenge in formulating the prefetching problem lies in effect-ively contextualizing accessed data and workloads for the prefetcher defining its output to enable accurate access prediction. Consider a query $q$ that accesses blocks $res^B_{q} \subseteq B$, where $B$ is the set of all data blocks in the database. Assuming $q$ requests $n$ blocks (i.e., $|res^B_{q}| = n$) in a specific order, the LBA sequence of these blocks can be represented as $res^{lba}_q = \langle lba_1, lba_2, \dots, lba_n \rangle$. The following outlines address-based and semantic-based prefetching formulations.
        }

    \vspace{-0.3em}  
        \subsubsection{Address-based prefetching}\label{sec:prelim_lba}

            \new{These prefetchers use sequences of LBA values or their deltas from prior queries. The delta sequence of $q$, denoted $res^{\Delta}_{q} = \langle ld_1, \dots, ld_{n-1} \rangle$, is computed by taking the difference between each consecutive LBA, as defined in Equation~\ref{eq:lba_conse_delta}. Address-based prefetching involves predicting the next LBA sequence $res^{lba}_{q_{n+1}}$ directly from $res^{lba}_{q_n}$ or via its delta sequence $res^{\Delta}_{q_n}$.}
            \vspace{-0.3em}  
            \begin{equation}\label{eq:lba_conse_delta}
              ld_i = lba_{i+1} - lba_i
            \end{equation}

        \subsubsection{Semantic-based prefetching}\label{sec:prelim_semantic}

          \new{Rather than relying on block addresses, semantic prefetchers predict $res^{lba}_{q_{n+1}}$ by leveraging information from block contents, either via semantic similarity~\cite{battle2016prefetching} or machine learning techniques~\cite{selep}. The SOTA prefetcher SeLeP~\cite{selep} treats block values as matrices and encodes them using AutoEncoder-based~\cite{autoencoder} feature extraction models.}

            \new{During preprocessing, non-numeric values are converted into text embeddings using Word2Vec~\cite{Word2Vec}, column values are normalized to reduce scale variance, and dimensionality is reduced through Principal Component Analysis (PCA)~\cite{pearson1901liii}.}

            \new{The query encoding for $q$ is generated by aggregating the encodings of the blocks in $res^B_{q}$. Due to the distinct semantics across different tables, the query encoding is structured as a matrix, where each row represents the aggregated block encodings for a specific table. SeLeP uses a sequence of recent query encodings to predict the next set of accessed LBAs.}

    \subsection{Problem Definition}\label{sec:problem_formulation}

    \new{In {\far}, the workload context is defined using both delta and semantic information from previous queries. Let $Cnx_{q_i}$ denote the context of query $q_i$, where $res^B_{q_i}, res^{\Delta}_{q_i} \in Cnx_{q_i}$. Accordingly, the address-based semantic prefetching problem is defined as follows: \textit{Given the contexts of the $l$ most recently executed queries, $\langle Cnx_{q_i} \rangle_{i=n-l}^n$, find and fetch the subsequent block access request, $res^{lba}_{q_{n+1}}$, by predicting $res^{\Delta}_{q_{n+1}}$.}}
    
    \vspace{0.25em}

      \new{By modeling each delta value as a class, a classification model can estimate the probability of each delta appearing in $res^{\Delta}_{q_{n+1}}$. Due to the wide range of both positive and negative deltas in large databases, regression is impractical, and classification results in a large output space. To address this, prefetchers restrict predictions to the top-$k$ most frequent deltas and map rare ones to a default class, skipping prefetching when only the default class is selected.}

    \subsection{Existing Prefetchers and Limitations} \label{sec:existing_prefetcher} 
        \subsubsection{{Traditional prefetchers}}  \new{These heuristics rely on block locality, prefetching sequential blocks, initiating prefetches after adjacent accesses~\cite{smith1978lookahead, opdenacker2007readahead}, or repeating recurring delta patterns~\cite{ki2000stride, naive}. Although effective for linear patterns such as full table scans, they perform poorly on irregular or random workloads.}
        
        \vspace{-0.35em}
        \subsubsection{{Learned prefetchers.}} 
      \new{By applying learning to historical access sequences, these prefetchers significantly improve performance, especially for irregular or complex workloads. Some use LBAs to model access patterns, while others incorporate data semantics.}
        
        \textit{Address-based prefetcher} proposed in \cite{chen2021revisiting} uses learning models to predict the next LBA via a two-level hierarchical structure, where each level is predicted separately. However, assigning a class to every possible LBA is inefficient for large and dynamic databases, as it results in excessive label space growth and limits scalability.
        
        \new{\textit{Semantic prefetchers} outperform address-based methods~\cite{tauheed2012scout, battle2016prefetching, selep}. SeLeP~\cite{selep} captures data semantics through offline block encoding (\S\ref{sec:prelim_semantic}). It dynamically clusters blocks into partitions based on recent co-access ratios and uses sequences of query encodings to estimate partition access probabilities. However, SeLeP faces scalability issues similar to direct LBA predictors. It also requires timely preprocessing of all blocks and retraining to incorporate new data into its predictions, limiting its suitability for workloads with frequent updates or bulk inserts where interactivity is critical.}

        \vspace{-0.35em}
        \subsubsection{{Memory prefetchers}} \new{Recent memory prefetchers have addressed scalability by estimating deltas instead of LBAs~\cite{2020delta_lstm, 2023sgdp}. These models typically forecast either a single delta or a fixed-length sequence of deltas. Predicting longer sequences requires repeated recursive inference or complex one-step models, both of which incur high computational costs. These approaches are effective when consecutive data accesses are localized to small memory regions with low delta diversity, but their performance deteriorates as the data space and delta variability increase.}

    \vspace{-0.6em}
  \subsection{Query Representation}\label{sec:query_encoding}
\vspace{-0.1em}

\new{Query statements and execution plans are key inputs for database tuning tasks such as index tuning~\cite{dbabandit, dbaregret,  jain2018query2vec, indexselect_AIMeetsAI}, view selection~\cite{hmab, viewselect_AVGDL}, and query optimization~\cite{qopt_bao, qopt_neo}. These systems encode query details into analyzable formats using lightweight or advanced techniques.}

\new{Lightweight methods encode query details such as accessed tables, query types, and normalized operation costs without relying on execution plans. In contrast, advanced techniques represent query plans to capture operation hierarchies. For example, Query2vec~\cite{jain2018query2vec} treats plans or SQL statements as sentences, strips literals and numbers, and encodes them using Doc2Vec~\cite{doc2vec}.}

\vspace{-0.6em}
\section{Motivation and Challenges}\label{sec:motivation}
    
    \new{Inspired by the success of delta modeling in memory prefetching, we explore its applicability to database semantic prefetching. Assessing this requires addressing two challenges: \textit{(i) How to define a unique LBA for blocks in a database? (ii) How to calculate $res^{\Delta}_q$ for a query $q$ when it accesses multiple blocks simultaneously?} We first explain how databases execute prefetching decisions.}

    \new{Prefetchers request candidate blocks from the database server, which locates them using unique internal identifiers distinct from storage-level LBAs. For instance, PostgreSQL uses CTID assigned to rows, Oracle employs a more detailed version called RowID, and Microsoft SQL Server uses RID. Prefetch decisions must be in the form of these internal pointers to allow the database engine to quickly complete the prefetching process.}

    \vspace{-0.75em}
    \subsection{\textit{Challenge (i)} — LBA Definition}
  \new{{\far} is deployed on PostgreSQL, where the CTID includes a block number and a row position within that block. Since I/O operates at the block level, only the block number is considered. However, CTIDs are unique only within a tablespace and may overlap across different tables, making raw CTIDs unsuitable as LBAs. This raises the challenge of defining unique LBAs from CTID values.}

    \begin{figure*}
  \centering
  \begin{minipage}[b]{0.41\textwidth}
    \centering
    \includegraphics[width=0.92\linewidth]{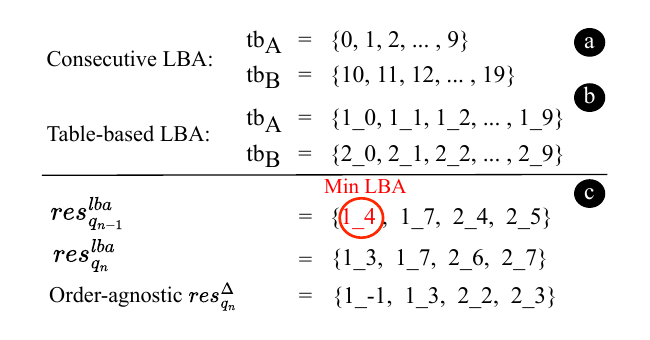}
    \vspace{-1em}
    \captionof{figure}{Examples for (a) consecutive LBA, (b) table-based LBA, and (c) order-agnostic delta calculation, based on min(LBA) of previously accessed blocks.}
    \label{fig:lba_eg}
  \end{minipage}\hfill
  \begin{minipage}[b]{0.56\textwidth}
    \centering
    \includegraphics[width=0.92\linewidth]{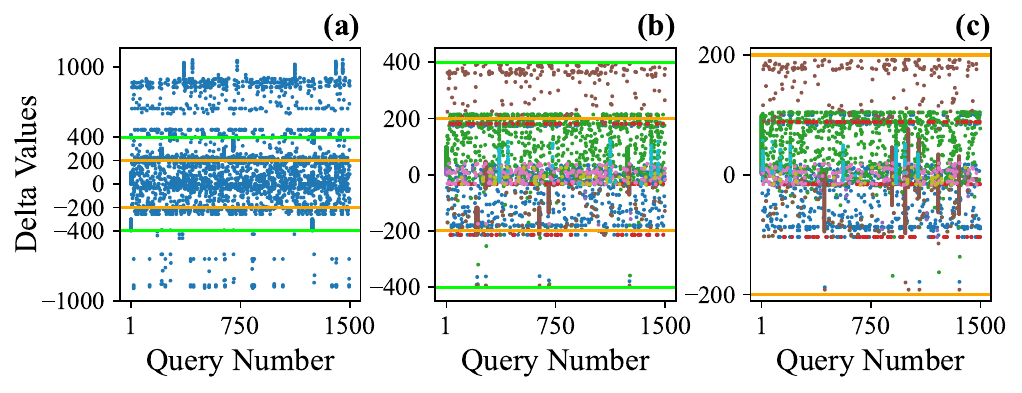}
    \vspace{-1.2em}
    \captionof{figure}{Delta values of Auction dataset with (a) consecutive LBA on SF=50, and (b-c) table-based LBA colored based on the tables on (b) SF=50 and (c) SF=25. Colored bands are added to assist readability.}
    \label{fig:deltas}
  \end{minipage}

    \vspace{-1.3em}

\end{figure*}

      \new{Memory systems often use consecutive hierarchical LBAs. This approach is similar to treating all database blocks as part of a single large table with unique CTIDs. However, this setup is unstable: inserting new blocks shifts existing LBAs and affects delta patterns. Moreover, queries often access blocks across multiple tables, producing large positive or negative deltas when switching tables.}

       \new{To address these issues, we implement a table-based LBA scheme with a two-level hierarchy. The first level represents the block’s CTID within its table, and the second identifies the table by its assigned table ID. Deltas are computed as hierarchical values, comprising the table ID and the difference in CTID values. For example, the delta from $b_i$ to $b_j$ with LBAs $tb_x\_CTID_{b_i}$ and $tb_y\_CTID_{b_j}$ is $tb_y\_(CTID_{b_j}-CTID_{b_i})$, where the table ID indicates the target table after applying the delta. Figure~\ref{fig:lba_eg}(a) and (b) show consecutive and table-based labeling on two sample tables, each with ten blocks.}

    \vspace{-0.75em}
    \subsection{\textit{Challenge (ii)} — Delta Calculation}
        \new{Evaluating LBA definitions requires a method to compute $res^{\Delta}_{q}$. In memory prefetchers, CPU instructions typically access a single data page, yielding a clear access order for delta calculation (Equation \ref{eq:lba_conse_delta}). In contrast, database queries often access \textit{sets} of zero to thousands of blocks, where defining a specific order is impractical~\cite{selep}.}
        
        \new{While we can sort accessed blocks by their LBA values and apply Equation \ref{eq:lba_conse_delta} to compute deltas, this approach has drawbacks. It requires multiple sequential predictions to generate prefetch decisions, introducing high latency that is incompatible with interactive workloads. In addition, prediction errors can propagate, degrading the accuracy of subsequent predictions.}

       To handle ordering and enable collective delta prediction, we compute deltas by subtracting each LBA in $res^{lba}_{q_i}$ from a reference LBA in $res^{lba}_{q_{i-1}}$. We evaluate three strategies for selecting the reference LBA—maximum, minimum, and median of the sorted LBAs—and measure their impact on the number of unique deltas and the hit ratio across three datasets.\footnote{Datasets and metrics are described in \S\ref{sec:datasets} and \S\ref{sec:metrics}, respectively.} As shown in Table~\ref{tab:reference_lba}, using $min_{{lba}}$ yields fewer unique deltas and better prediction accuracy. Hence, we adopt $min_{lba}$ and compute the order-agnostic delta set using Equation \ref{eq:order_agnostic_delta}. Figure \ref{fig:lba_eg}(c) shows an example delta set for a sample query using the min table-based LBA, highlighted in red.
    \vspace{-0.25em}
        \begin{equation}\label{eq:order_agnostic_delta}
              {res}^{{\Delta}}_{q_i} = \left\{ \text{LBA} - \min\left({res}^{{lba}}_{q_{i-1}}\right) \, \middle| \, \text{LBA} \in {res}^{{lba}}_{q_i} \right\}
    \vspace{-0.75em}
        \end{equation}
        

        \begin{table}[htbp]
          \centering
          \caption{{Total Number of Unique Deltas and Hit Ratio with Different Labeling Methods \small{(best in bold, second best underlined)}}}
          \vspace{-1.15em}
          \scalebox{0.82}{
          \begin{tabular}{|c|ccc|ccc|}
            \hline
            Dataset
            & \multicolumn{3}{c|}{Hit Rate (\%)}
            & \multicolumn{3}{c|}{Delta Count} \\
            \hline
            
            Reference LBA& Min    & Median & Max    & Min     & Median & Max     \\
            \hline
            TPC-C   
              & \textbf{96.7} & 94.86      & \underline{94.94}
              & \underline{5511} & 5711     & \textbf{5508} \\
            Auction 
              & \textbf{94.46} & 91.02      & \underline{92.27}
              & \underline{2263} & \textbf{2132} & 2483      \\
            SDSS    
              & \textbf{91.93} & 91.03      & \underline{91.56}
              & \textbf{13992} & 18500    & \underline{18041} \\
            \hline
          \end{tabular}
          }
          \vspace{-1.7em}
          \label{tab:reference_lba}
    \end{table}

    \subsection{Delta Analysis}\label{sec:delta_analysis}
        Figure \ref{fig:deltas} shows the delta values for 1500 queries from datasets with different scale factors (SF, indicating dataset size) in the Auction benchmark, calculated using Equation \ref{eq:order_agnostic_delta} and the two labeling methods. Comparing Figures \ref{fig:deltas}(a) and \ref{fig:deltas}(b) reveals that consecutive LBAs result in a much wider delta range, while table-based labeling reduces it by over 60\% on the same data and workload.

      \new{Analyzing the delta values reveals several key patterns. First, most deltas fall within a specific, bounded range and are highly concentrated around zero, indicating frequent reuse of a limited set of delta values. This suggests the viability of predicting block accesses using delta modeling approach.}
    
     \new{Second, delta visualizations reveal noticeable patterns in delta occurrences, especially in Figures~\ref{fig:deltas}(b) and (c), where deltas are color-coded by table ID. These patterns suggest that deltas are not random and can be modeled for prediction.}
    
       \new{Third, we analyze deltas across datasets with varying SFs from the same benchmark. Figures~\ref{fig:deltas}(b) and (c) show results for SF=50 and SF=25 of the Auction benchmark. Delta patterns remain consistent across datasets with similar schemas and workloads, though higher SFs may trigger different query plans, subtly altering access behavior. As SF increases, delta range and density both expand, with non-linear changes in delta values and per-query counts. Larger datasets exhibit broader ranges and more concentrated distributions, indicating that delta patterns scale with dataset size while preserving underlying structure. This inspired our design of a prefetcher that generalizes patterns learned on smaller datasets to larger ones.}

    \begin{figure*}
        \centerline{\includegraphics[width=\textwidth]{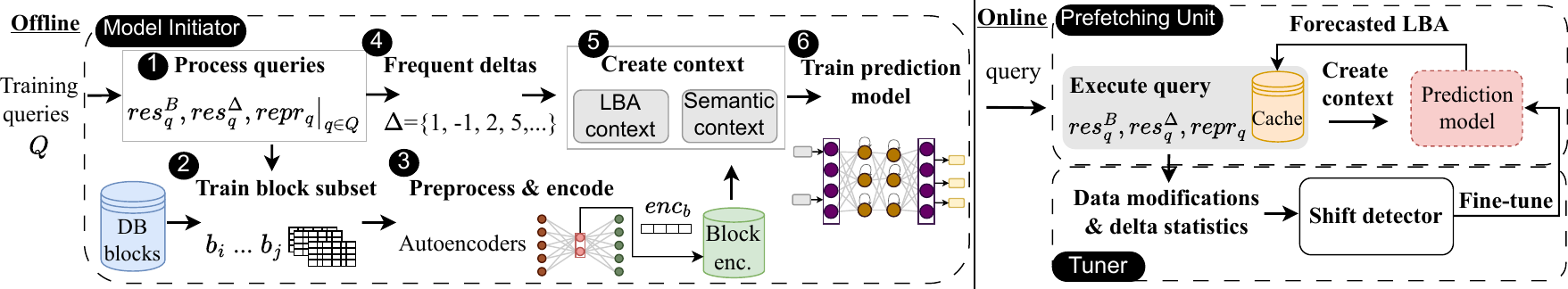}}
        \vspace{-1em}
        \caption{{System architecture of {\far}}}
        \vspace{-1.2em}
        \label{fig:model_overall}
    \end{figure*}

    \vspace{-0.5em}
\section{ {\far} Framework}

\new{{\far} is a learning-based framework that models delta patterns by leveraging semantic relationships extracted from previously accessed data, executed queries, and contextual features. Given PostgreSQL’s block size of 8kB to 32kB, a large dataset can span millions of blocks, resulting in a vast address space and diverse delta values, complicating accurate prediction. To mitigate this, we logically group $lb_{size}$ sequential blocks for prediction purposes, while caching remains at the native block level.\footnote{The impact of $lb_{size}$ is evaluated in~\S\ref{sec:sensitivity}.}}

Accurate delta prediction requires transforming contextual information into representations that are both suitable for processing and adaptable to change. {\far} achieves this through several key internal components, described in the following sections.

    \vspace{-0.6em}
\subsection{An Overview}

\new{{\far} consists of three components. The \textbf{model initiator} sets up statistical and learning-based models and the frequent delta set. The \textbf{prediction unit} prepares contexts and generates prefetch decisions using the trained models. The \textbf{tuner} continuously refines the models to adapt to evolving data and workloads. Figure~\ref{fig:model_overall} illustrates {\far}'s architecture, and Table~\ref{tab:notations} lists frequently used notations.}

    The model initiator processes training queries, collects their results, and selects the accessed blocks along with their immediate neighbors for semantic extraction. For each table, its blocks are preprocessed and input into an autoencoder (Figure \ref{fig:model_overall}-1, 2, and 3) to produce block encodings for semantic contexts.

    The semantic context generator creates query semantics by combining the query result encoding ($enc^B_q$), aggregated from encoding of its accessed blocks, with the query statement representation ($repr_q$). Using a plan-agnostic approach, {\far} encodes features extracted directly from the query statement. This process outputs query semantics that serve as input for the prediction model.

    To establish the LBA context and select the prediction model’s delta classes, {\far} analyzes deltas within the training workload and limits their count to the most frequent ones. It generates a binary representation of query deltas ($bi\_delta_q$) and a one-hot encoding based on the size of $bi\_delta_q$, which serve as the model's LBA-based input context. By integrating sequences of semantic and LBA contexts with the last accessed table, {\far} trains its prediction model to learn delta patterns (Figure \ref{fig:model_overall}-4, 5, and 6).

    The prediction unit uses the established system to construct context sequences and make prefetch decisions. It predicts the next accessed tables, a probability distribution over delta classes, and an estimated delta count \textit{n}, from which the top-\textit{n} deltas are selected. The predicted tables and deltas are then combined to form table-based deltas. It then filters these deltas based on their frequency in the historical workload and selects candidate LBAs for prefetching.

    The tuner maintains system adaptability by responding to shifts in data and workloads. Its responsibilities include updating block preprocessing components, revising frequent historical deltas, and tuning the autoencoders and prediction model. Also, the tuner encodes newly inserted blocks and those accessed for the first time.

\begin{table}[htbp]
        \caption{{Frequently Used Notations}}
        \vspace{-1.3em}
        
        \begin{center}
                \scalebox{0.84}{
        \begin{tabular}{|c|c|}
            \hline
            {Symbol}&{Definition}\\
        	\hline
            $res^B_q$ & Set of block data accessed by query $q$ \\
            \hline
            $res^{lba}_q$, $res^{\Delta}_q$ &\centered{Set of LBAs of $q$ result, $q$ result deltas} \\ \hline
            $enc^B_q$, $repr_q$ & $q$ result encoding, $q$ statement representation \\ \hline
            $bi\_delta_q$ & binary representation of $res^{\Delta}_q$\\ \hline
            $\Delta$ & Frequent deltas selected by the model where $|\Delta|=ds$ \\ \hline
            $\tau$, table $\alpha$ & table selection threshold and its modifying factor \\ \hline
            $k_{dc}$ & Multiplier for predicted query delta count \\ \hline 
            $lookback$ &\centered{Context sequence length used by the model} \\
            \hline
            $k$ & Prefetch size in unit of 128 blocks \\ \hline 
            
        \end{tabular}
         }
         \label{tab:notations}
        \end{center}
      
    \end{table}

    \vspace{-0.8em}
    \subsection{Block Encoding}
        Semantic prefetchers leverage the actual data values to make prefetch decision. Since each block can contain hundreds of values, these prefetchers need to create a concise representation for the blocks which summarizes their key characteristics. 
        
        Since the data stored in a database can be used for various purposes, it is impractical to determine which attributes hold the most critical information. Thus, block semantics are extracted using unsupervised feature extraction methods such as Autoencoders \cite{autoencoder}.
        
        \new{We enhance the block encoding component of SeLeP \cite{selep}, described in \S\ref{sec:prelim_semantic}. This component processes and encodes each table’s blocks into a compact representation using a table-specific autoencoder. The autoencoders, implemented as multilayer perceptrons (MLPs), are table-specific because differences in schema, size, and semantics make a shared model ineffective.}
        
        Before applying statistical and learning methods to block data, non-numerical values must be converted into numerical representations. The Word2Vec encoding approach that is used by SeLeP cannot handle unseen data, leading us to evaluate two alternatives: FastText \cite{bojanowski2017fasttext}, which extends Word2Vec by learning embeddings for strings and their substrings, and MinHash \cite{broder1997minhash}, which is a Locality-Sensitive Hashing (LSH) technique capable of encoding strings.

        Employing these methods resulted in much higher encoding times than Word2Vec, with FastText requiring orders of magnitude longer training. In contrast, Word2Vec supports incremental updates, where new words refine the existing embedding space instead of rebuilding it. Following the iterative retraining strategy explored in~\cite{dynamicWord2vec, incrementalWord, dynamicword2}, we retain Word2Vec with an added mechanism to dynamically expand its vocabulary. Textual values from the block are combined into a sentence and fed into the model, enabling it to learn embeddings for new values as they are encountered.

        The most effective part of data preprocessing is data normalization as training the Autoencoders on the raw data blocks with wide range of values will result in a poor block encoding. We retain the min-max normalization method (Equation \ref{min-max_eq}) from~\cite{selep} since, in a dynamic dataset, maintaining minimum and maximum values is simpler and more computationally efficient compared to other statistical metrics, such as mean, standard deviation, or quartiles, used by alternative normalization methods.
        \vspace{-0.1em}
        \begin{equation}
             x_{normalized} = \frac{x-min(X)}{max(X) - min(X)}\times 2-1\label{min-max_eq}
        \end{equation}

        To address tables with a large number of columns,~\cite{selep} applies PCA to the data after normalization. In {\far}, this step is made more adaptive by replacing PCA with Incremental PCA (IPCA) \cite{Bishop2007IPCA}, which efficiently updates the transformation as new data is added, \emph{eliminating the need for a complete re-computation} \cite{halpern2018advances}.

        Once the block dimensions are reduced, the processed data is fed into an autoencoder corresponding to its table to generate encodings. These encodings are stored for later use in creating query result encodings. In \S \ref{sec:tuner} we explain how IPCA is leveraged to evaluate whether tuning the autoencoders is necessary.

    \vspace{-0.4em}
    \subsection{Context Creator}
        {\far} leverages both semantic and LBA-based contexts for access prediction. The semantic context captures the meaning and structure of recent queries, while the LBA-based context encodes information about the delta values associated with those queries. This section details the generation of these contexts.
        
   \vspace{-0.7em} 
    \subsubsection{Query semantics generator}

        {A query’s behavior depends not only on the blocks it accesses but also on how it filters, joins, and aggregates data—details that block-level embeddings alone cannot capture. Query statements more fully express user intent and provide richer context, enabling more accurate prediction of future data accesses within a query session (a series of closely timed, goal-aligned queries). This is particularly important in exploratory workloads, where sessions aim to uncover specific insights.}
        
        {Incorporating statement representations strengthens table access modeling by embedding table interactions directly into the query semantics.  It also reduces reliance on query result encodings, which may degrade when underlying data changes. Hence, {\far} combines query's result encoding ($enc^B_q$) and statement representation ($repr_q$) to create a robust query semantic encoding ($enc_q$).}

\begin{figure}
           \centerline{\includegraphics[width=.95\linewidth]{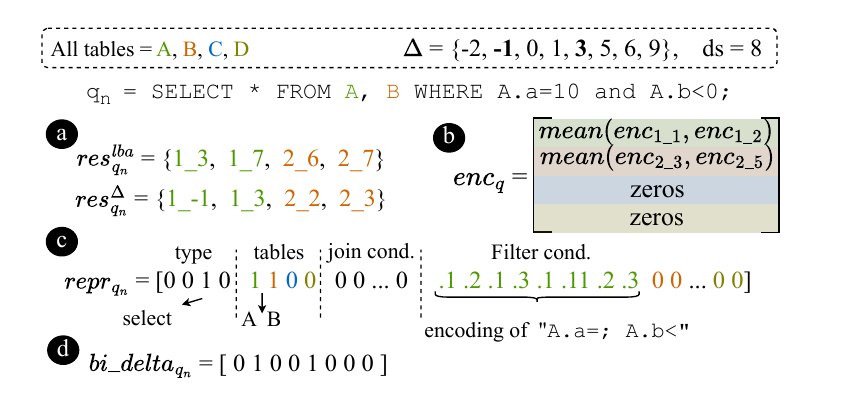}}
           \vspace{-1em}
            \caption{Examples of (a) result LBAs and \textit{delta}s, (b) result encoding, (c) statement representation, and (d) binary delta.}
            \vspace{-1.3em}
            \label{fig:q_semantic}
        \end{figure}

        \vspace{-0.5em}
        \paragraph{\textbf{Query result encoding}} $enc^B_q$ can be calculated by aggregating $enc_b$ of the blocks in $res^B_{q}$. However, aggregating encoding of blocks from different tables will result in a meaningless representation \cite{selep}, since the semantic interpretation of individual fields within $enc_b$ varies across tables. Thus, the query result must be encoded as a matrix where encodings of blocks from the same table are aggregated and placed in a single row corresponding to that table. Figure \ref{fig:q_semantic}(b) depicts $enc^B_q$ for a sample query accessing n blocks.

        \vspace{-0.4em}
        \paragraph{\textbf{Query statement representation}} Constructing an effective statement representation requires addressing three key questions:  \textit{(i) Which query details are most relevant to the task? (ii) Is including additional information from the query plan beneficial? (iii) Should the representation maintain consistent semantic meaning across queries?}

        Different systems selectively encode details tailored to their specific tasks. For instance, QTune \cite{li2019qtune} encodes accessed tables and operation costs, while {\small{DBABandit}} \cite{dbabandit} focuses on accessed columns only. In prefetching, the emphasis is on data accessed by a query, which depend on its type, accessed tables, join conditions, accessed columns, and filters. Different query types exhibit distinct block access patterns: modification queries usually access fewer blocks within a single table, whereas selection queries often join multiple tables and access more blocks. Join and filter conditions narrow the query's target, dictating which specific blocks must be read.

        {We define a compact, structured representation of the statement capturing these details. The query type is encoded as a one-hot vector, and since queries may access multiple tables, we represent them with a binary bitmap, setting bits for each referenced table.}

        {Encoding query conditions is more challenging, as multiple conditions can apply to any column within a table. We parse the query execution plan and extract join and filter predicates of each table. Following Query2Vec~\cite{jain2018query2vec}, we remove numeric values and literals from the conditions to improve generalizability. Each table’s conditions are then treated as a short document and encoded using a Doc2Vec~\cite{doc2vec} model. We have separated the join conditions with the filters since they have distinct impact on the accessed blocks.}

        {The final statement representation is composed of 4 bits for the query type, $|TB|$ bits for accessed tables, and two parts of $8\times |TB|$ bits each capturing encoded joining and filtering conditions applied to each table. This format ensures uniform representations where each field has a consistent and comparable meaning across queries. Figure \ref{fig:q_semantic}(c) shows $repr_q$ for $q$ with filter conditions on table A.}

        {We focus on table-level details because including every column from all tables in the database results in a high-dimensional query representation with many zeros, as most queries access only a small subset of columns. We deliberately exclude lower-level plan details such as join strategies or operation order, as they introduce unnecessary specificity and reduce generalizability.}

        {We evaluate the impact of our query representation by comparing {\far} to alternative variants that use different encodings: no query representation (None), {\small{SQL-Query2vec}} (Sq2v), {\small{Plan-Query2vec}} (Pq2v)} as detailed in \S \ref{sec:query_encoding}, and a partial representation that includes query type and accessed tables only (Simple). We assess the prefetcher's hit ratio and average recall across four test datasets\footnote{The datasets are described in \S \ref{sec:datasets} and the metrics are provided in \S \ref{sec:metrics}}. 

        \begin{table}[htbp]
        \caption{ Hit Ratio, Average Recall and Average Query Statement Encoding Time of Different Representation Methods}
        \vspace{-1.3em}
        \begin{center}
                \scalebox{0.81}{
        \begin{tabular}{|*6{c|}c|}
        	\hline
        	\multicolumn{2}{|c|}{Test}&None&Sq2v&Pq2v&Simple&\far\\
        	
            \hline	\multirow{2}{*}{{Auction}}
        	&\small{Hit Rate}&94.11&94.42&\underline{94.66}&94.15&\textbf{95.26}\\
        	 &Recall&57.94&71&71.37&\underline{71.63}&\textbf{72.01}\\
             
        	\hline	\multirow{2}{*}{\shortstack{Auction 20\% \\ size ratio}}
        	&\small{Hit Rate}&91.22&92.37&92.75&\underline{96.11}&\textbf{96.58}\\
        	 &Recall&54.56&58.79&58.53&\underline{59}&\textbf{60.01}\\
        	
            \hline	\multirow{2}{*}{{SDSS}}
        	&\small{Hit Rate}&97.89&\textbf{98.56}&97.6&\underline{98.1}&97.91\\
        	 &Recall&73.41&72.87&72.45&\textbf{75.2}&\underline{74.12}\\
            
            \hline	\multirow{2}{*}{\shortstack{SDSS 10\% \\ size ratio}}
        	&\small{Hit Rate}&97.89&\underline{98.08}&97.51&\textbf{98.37}&\textbf{98.37}\\
        	 &Recall&72.56&71.83&70.63&\underline{75.52}&\textbf{76.57}\\
        	\hline
        	\multicolumn{2}{|c|}{Avg preparation time/$q (ms)$}&NA&2.72&2.69&\textbf{0.45}&\underline{0.52}\\
        	\hline
        \end{tabular}
     }
     \vspace{-1.5em}
     \label{tb:query_representation}
    \end{center}
    \end{table}

        The results in Table \ref{tb:query_representation} show that our plan-agnostic method (\far) and its partial version (Simple) achieve the best performance, especially compared to None. In datasets with size ratio, where the prefetcher is trained on a sampled dataset and tested on a larger one, {\far} exhibits the greatest improvement over None.

    \vspace{-0.3em}
    \subsubsection{LBA-context creator}\label{sec:lba_context_generator}
        The LBA-based context is derived from $res^{\Delta}_q$ of recent queries. Large databases can have vast number of  possible delta values, increasing model complexity and reducing accuracy if all are included.
        To address this, we must select a subset of these deltas to define the model output and the LBA-context.
        
        {\far} predicts future accesses by modeling deltas, making the choice of delta values central to its design, as they directly influence input features, output classes, and overall model complexity. Since most queries exhibit a small set of frequent deltas (\S\ref{sec:delta_analysis}),  selecting a subset of these frequent values simplifies the prediction process and ensures the model focuses on the most impactful values.

   \new{To identify effective deltas ($\Delta$), we analyze $res^{\Delta}_q$ from the training workload, discard table identifiers, and retain unique CTID deltas, denoted as $delta_q$. We compute delta frequencies and select the top $ds$ most frequent values for the prediction model (Figure~\ref{fig:model_overall}-4). The impact of $ds$ values is evaluated in \S~\ref{sec:sensitivity}, with 1500 chosen as the optimal setting, which performs well on a 155GB database.}
   
        {To handle infrequent deltas not included in $\Delta$, we introduce a default class to ensure full coverage. $delta_q$ is then} encoded as a binary vector, $bi\_delta_q$, of length $ds$, where $bi\_delta_q[i] = 1$ if $\Delta[i] \in delta_q$. If no value of $delta_q$ is in $\Delta$, {only the default class is set.} Figure~\ref{fig:q_semantic}(d) shows $bi\_delta_q$ of a sample query with $ds=8$. 
        
        $bi\_delta_q$ describes recent delta patterns to the model, computed relative to a $min_{lba}$ value using Equation \ref{eq:order_agnostic_delta}. To enrich the LBA-based context and enhance delta modeling, we also encode $|{delta_q}|$ as a one-hot vector and include the table ID of the $min_{lba}$.

    \subsection{Delta Modeling}
        This section explains the details of {\far}'s prediction model. 

        \vspace{-0.1em}
        \subsubsection{Input} {\far} uses last $n_{lookback}$ query contexts ($cnx_q$), each combining LBA-based and semantic components, as input to capture temporal dependencies across queries. This approach accounts for the impact of one query’s results on formation of subsequent queries \cite{StratosExploration, ResearcherGuide}. Impact of $n_{lookback}$ values is evaluated in \S \ref{sec:sensitivity}.

        \vspace{-0.1em}
        \subsubsection{Output} 

        \new{The multi-task prediction model forecasts} three aspects of the next query $q_{i+1}$: the accessed tables, the delta classes for values in $delta_{q_{i+1}}$, and the count of deltas in $delta_{q_{i+1}}$. These predictions are combined to generate table-based LBAs to be prefetched.

        \begin{figure}
            \centerline{\includegraphics[width=\linewidth]{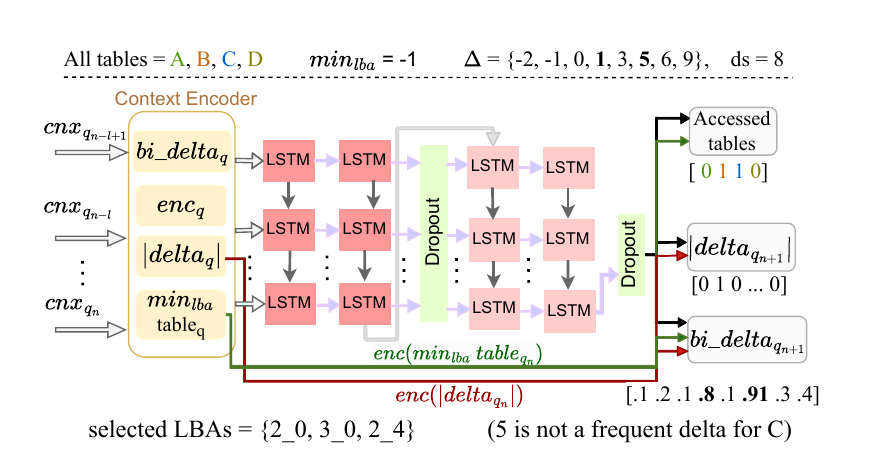}}
            \vspace{-0.7em}
            \caption{The prediction model architecture and a sample prediction output. Context components are encoded separately and merged to be fed in LSTM layers.}
            \vspace{-1.75em}
            \label{fig:prediction_model}
        \end{figure}
        
        \vspace{-0.1em}
        \paragraph{\textbf{Accessed tables}} Query accessed table is represented with a bitmap vector where each field corresponds to a specific table and value 1 indicates the query accesses blocks within that table. {\far} estimates the probability of each table being accessed in $q_{i+1}$, and converts these probabilities into binary values using a threshold $\tau$.
        
      \new{Since workloads are dynamic, a static threshold is unsuitable, so {\far} updates $\tau$ at each step using the minimum predicted probability of accessed tables and the count of false negatives, as defined in Equation~\ref{eq:tb_thresh} with $\alpha=0.05$ (evaluated in \S\ref{sec:sensitivity}). The adaptation aims to maximize recall, ensuring all positive tables are predicted.}
        
        \vspace{-0.6em}
        \begin{equation} \scalebox{0.94}{$
                \tau=\begin{cases}\tau - \alpha \times \text{\small{|False Negatives|}}, & \text{if } \tau < \min(\text{P\small{(Accessed Tables)}}) \\ 
                \tau+ \alpha/10, & \text{otherwise} \end{cases}\
                 $}
        \label{eq:tb_thresh}  
        \end{equation}
        
        \vspace{-0.7em}

        {\paragraph{\textbf{Delta count}} SOTA prefetchers prefetch a fixed number of blocks after each access, such as 9 blocks per access in~\cite{chen2021revisiting} or 40 partitions (128 blocks each) per query in~\cite{selep}. In contrast, {\far} predicts the delta count ($|delta_{q}|$) for the next query and dynamically adjusts the prefetch size to better accommodate variable query result sizes.}
        
        Accurately predicting $ |delta_{q}| $, is challenging as it depends on factors such as data distribution and query predicates
        . Moreover, since delta predictions are not perfectly precise, some extra blocks must be prefetched to ensure a high performance. To address this, {\far} scales the predicted delta count by a factor $k_{dc}$ and prefetches $ |\hat{delta}_{q}| \times k_{dc} $ blocks. The impact of $ k_{dc} $ is evaluated in \S \ref{sec:sensitivity}.

        \vspace{-0.3em}
        \paragraph{\textbf{Delta values}} \new{The final output predicts the likelihood of each delta class in $bi\_delta_{q_{i+1}}$. The predicted deltas are combined with the predicted tables to generate candidate table-based deltas. However, using all such combinations is inefficient, and complex filtering is impractical within the limited time before the next query. To address this, {\far} filters deltas based on their occurrence frequency in system history. It maintains a per-table lookup of frequent deltas, dynamically updated with recent queries, ensuring that only commonly observed deltas are used to construct LBAs for prefetching.}

        \vspace{-0.3em}
        \subsubsection{Model Architecture} To capture temporal dependencies in delta patterns, {\far} employs an LSTM-based architecture. LSTM networks are a type of Recurrent Neural Network (RNN) capable of learning sequential dependencies in data by maintaining an internal state over time. This makes them particularly suitable for modeling sequences in prediction tasks, including prefetching \cite{2020delta_lstm, selep}. 
        
        Although more complex models like Transformers~\cite{transformer1, transformer2} have become popular, {LSTMs remain a strong choice for systems requiring faster and simpler training and inference.} In addition to the LSTM-based model, we implemented and evaluated {\far} using two other models, a three-layer MLP network and a two-layer Convolutional Neural Networks (CNN). {Results show that the LSTM architecture achieves comparable or better recall while prefetching up to 18\% fewer blocks, demonstrating its ability to capture temporal patterns effectively for accurate and efficient prefetching.}

        {Figure \ref{fig:prediction_model} shows the prediction model. It receives $\langle cnx(q_{i})\rangle_{i=n-l}^n$ and compresses each context component separately, using time-distributed Dense layers. The resulting embeddings are concatenated into a sequence of 128-dimensional vectors and passed to the LSTM layers. The LSTM output summarizes the recent workload and is used as a shared representation for the Dense layers, each generating a certain output with additional task-specific inputs.}

        The delta count is produced by a Dense layer with Softmax activation, which additionally takes the encoding of the last $|delta_q|$ as input. Accessed tables and delta values are predicted by Dense layers with sigmoid activation that additionally incorporate the embedding of the last accessed table and $bi\_delta_q $, respectively.

        \vspace{-0.3em}
        \subsubsection{Training configuration}
        Binary cross-entropy (BCE) works well for multi-label classifications such as our prefetching problem since it treats each label as an independent binary problem, optimizing predictions for each label separately~\cite{chen2021revisiting, selep, 2023sgdp}. However, the prefetching classification task usually faces significant class imbalance that necessitates adjustments to this loss function.
        
        \vspace{-0.4em}
        \paragraph{Class imbalance challenge} In query workloads, the blocks are accessed unevenly. Our delta analysis (\S \ref{sec:delta_analysis}) reveals that even among the frequent deltas, certain values occur more often. This leads to an uneven class distribution, where majority classes are overrepresented and minority classes are sparse. Predicting frequent classes becomes easy, while minority classes are treated as harder cases.

        \begin{figure}
            \centerline{\includegraphics[width=0.8\linewidth]{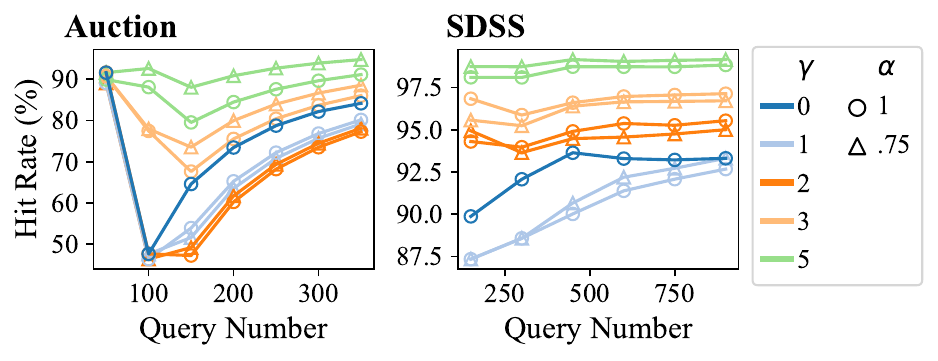}}
            \vspace{-1.5em}
            \caption{$FL$ setup impact on hit rate of {\small{100}}-query windows.}
            \vspace{-1.8em}
            \label{fig:loss}
    \end{figure}

        To address this imbalance, we employ Focal Binary Cross-Entropy Loss ($FL$), which extends standard BCE by adding a modulating factor that down-weights easy examples and focuses training on hard, misclassified ones. The $FL$ loss is defined as:
        \begin{equation}
                FL(\hat{y}) = -\alpha(1-\hat{y})^{\gamma}log(\hat{y})
        \label{eq:focal_loss}            
        \end{equation}
        where $\hat{y}$ is the predicted probability for the true class, $\gamma>1$ reduces the loss contribution from easy cases, focusing on harder ones, and $\alpha$ adjusts the overall loss contribution of the class, with higher $\alpha$ increasing the weight of the minority class in the total loss.

        {Figure~\ref{fig:loss} presents the hit ratio under different $FL$ setup, where $\alpha{=}1, \gamma{=}0$ shows plain binary cross-entropy (BCE) loss.  Poor BCE performance  stresses the need for better handling of class imbalance. Increasing $\gamma$ improves hit ratio but also causes greater training and prediction instability. Although $\gamma{=}5$ has the highest hit ratio, we select $\gamma{=}3$ for a better trade-off between accuracy and stability.}

    \vspace{-0.5em}
        \paragraph{Overfitting Challenge} Imbalanced classes can cause overfitting by making the model overly biased toward the frequent classes, resulting in poor generalization for the minority classes. Additionally, using $FL$ loss, the model may overemphasize the minority classes and overfit to rare cases, reducing overall performance. To mitigate this, we incorporate dropout layers to regularize learning, use a low learning rate for stable convergence, and employ a large batch size to reduce gradient noise and enhance generalization.

    \subsection{Tune and Generalize} \label{sec:tuner}
        Regular system tuning is crucial in dynamic environments where both data and workloads evolve over time. {\far} adapts its components to accommodate these changes, ensuring stable performance.

        \vspace{-0.3em}    
        \subsubsection{Block encoding} \new{The IPCA in {\far} also identifies shifts in data distribution by comparing the cosine similarity of principal components (PCs) before and after fitting new data. Small data batches typically have minimal impact, with similarities often above 0.8. For similarities below 0.8, {\far} fine-tunes the table’s autoencoder using the new data and re-encodes only the new blocks, leaving previous encodings unchanged to save time. Query semantics help mitigate any inconsistencies and maintain performance.}

    \vspace{-0.3em}
        \subsubsection{Deltas} \new{Frequent deltas change with workload shifts. The tuning component tracks delta frequencies and updates the lookup tables. After $l_{tune}$ queries, {\far} refreshes $\Delta$ with the most recent frequent deltas. Since modifying $\Delta$ alters the model’s output, {\far} fine-tunes the model by freezing all layers except the final dense layers and retraining on the recent workload for 15 epochs with a low learning rate. The impact of $l_{tune}$ is analyzed in \S \ref{sec:sensitivity}.}

        \vspace{-0.3em}
        \subsubsection{Generalizability}
      \new{{\far}'s tuning capabilities allow it to be trained on a smaller dataset and deployed on a much larger one. It gradually adjusts the IPCA and autoencoders to the new data and updates the prediction model to handle new delta values. Larger databases typically have a wider range of frequent deltas (\S~\ref{sec:delta_analysis}), requiring a larger output size for the prediction model compared to the training dataset. To address this, {\far} includes $ds$ void classes, initially unassigned, during training so newly detected deltas can be mapped to these classes after deployment, enabling faster fine-tuning and better adaptation to the larger database.}

\vspace{-0.3em}
\section{Experiment Settings}
    We have evaluated {\far} across a wide range of real-world and benchmark datasets with analytical read-only workloads and hybrid analytical-transactional workloads. In this section we explain our experimental setting and the test databases used in the experiments.
    
    \vspace{-0.5em}
    \subsection{Implementation and Configurations}
        {{\far} is implemented in python using TensorFlow/Keras framework \cite{tensorflow2015-whitepaper}. LSTMs are configured with 64 cells and trained in batches of 128, using early stopping on delta prediction loss (validated on 10\% of the training data) or a maximum of 25 epochs. The prediction model is trained using $FL$ loss ($\alpha=0.75$, $\gamma=3$), while the autoencoders use mean squared error; both are optimized with Adam~\cite{kingma2014adam}, using learning rates of 0.0001 and 0.001, respectively.}
        
    \new{{\far} is deployed on PostgreSQL~\cite{stonebraker1986design}, using the pg\_prewarm module in buffer mode to fetch blocks by CTID as a background task. After each query, it selects candidate blocks, computes contiguous CTID ranges, and issues prewarm commands—terminating early if a new query arrives to avoid interfering with query I/O.}

         \begin{table}[htbp]
        \vspace{-0.4em}
            \caption{Datasets and Workload Summary}
            \vspace{-1em}
            \begin{center}
                \scalebox{.81}{
                \begin{tabular}{>{\raggedright\arraybackslash}p{2.05cm}>{\centering\arraybackslash}p{.55cm}>{\centering\arraybackslash}p{.85cm}>{\centering\arraybackslash}p{.5cm}>{\centering\arraybackslash}p{1.3cm}>{\centering\arraybackslash}p{.6cm}>{\centering\arraybackslash}p{.85cm}>{\centering\arraybackslash}p{.6cm}}
                     Property & {\rotatebox{0}{SDSS}} & {\rotatebox{0}{\small{Genomes}}} & {\rotatebox{0}{Birds}} & \small{\rotatebox{0}{{TPC-H Skew}}} & \small{\rotatebox{0}{TPC-C}} & {\rotatebox{0}{{Auction}}} & {\rotatebox{0}{{Wiki}}} \\ 
                    \hline
                    Scale factor & - & - & - & 30 & 250 & 50 & 100 \\
                    Size (GB)    & 155   & 10  & 8  & 70 & 25  & 16 & 21 \\
                    Tables & 95 & 13 & 6 & 8 & 9 & 16 & 9\\ 
                    {\small{Read-only queries }} & 100\% & 100\% & 100\% & 100\% & 8\% & 55\% & 92.2\% \\ [0.2em]
                    \small{Avg$(|res^{lba}_q|)$} &9.74& 20.7& 7.3& 29699 & 2.44& 42.9& 48.5\\ [0.1em]
                    \small{min$(|res^{lba}_q|)$} &1& 1& 1& 11 & 1& 1& 1\\ [0.1em]
                    \small{max$(|res^{lba}_q|)$} &643& 607& 201& 125000 & 108& 12500& 3171\\ [0.1em]
                    \small{Train workload} & 220k&10k&10k& 10k& 15k & 100k & 15k\\ 
                    \small{Test workload} & 1000&450&300& 50&500 &400 & 600\\ 
                    \hline
                \end{tabular}
                }
                \vspace{-1.4em}
            \label{tab:db_workload}
            \end{center}
        \end{table}

        Unless stated otherwise, the configuration of the experiments is: cache size = 8GB for datasets larger than 21GB and 4GB otherwise, logical block size = 32 blocks, block size = 32kB, $n_{lookback}$ = 2, delta size = 1500, initial $\tau$ = 0.1, table $\alpha$ = 0.1, and $k_{dc}$ = 25. All values are selected based on our sensitivity analysis presented in \S \ref{sec:sensitivity}.

        \textit{Hardware.} Experiments are conducted on an Ubuntu server equipped with 48 Core at 2.4GHz, 1.1TB RAM, 10K RPM disk, and one NVIDIA V100 GPU with 16GB memory. To isolate prefetching effect, the operating system cache is flushed after each query.
    
    \vspace{-0.3em}
    \subsection{Datasets and Workloads} \label{sec:datasets}
        Our evaluation datasets are summarized in Table \ref{tab:db_workload}.

        \vspace{-0.4em}
        \subsubsection{Analytical}
        Three real-world datasets are used for analytical tests: SDSS, Birds, and Genomes. \textbf{SDSS} is a subset of the seventh Data Release (DR7) of Sloan Digital Sky Survey \cite{abazajian2009sdss} extended from MyBestDR7~\cite{skyserver}
        using SciScript
        library~\cite{sciscript}.
        \textbf{Birds} and \textbf{Genomes} are datasets from the SQLShare project \cite{sqlshare}, containing primarily textual data on bird species and genomic information, respectively.
    
        \vspace{-0.4em}
        \subsubsection{Hybrid} We utilize Benchbase tool \cite{DifallahPCC13} to generate three benchmark datasets: \textbf{TPC-C}, AuctionMarket (\textbf{Auction}), and Wiki-pedia (\textbf{Wiki}). To get their test workloads, we run Benchbase with its default settings for 2 minutes and collect all executed queries. 

        \vspace{-0.4em}
        \subsubsection{Generalized}\label{sec:generalization_workload} To evaluate {\far}'s generalization capacity, we consider the databases in Table \ref{tab:db_workload} as the target database for testing and prepare various versions of a dataset with smaller sizes as a train dataset. The following train dataset sizes are used: 16GB and 90GB for SDSS, SF 1, 10, 50, 100, 150, 200 for TPC-C, SF 1, 10, 25 for Auction, and SF 1, 10, 25, 50 for Wiki.

        \vspace{-0.4em}
        \subsubsection{Skewed}\label{sec:skewed_workload} For completeness, we evaluate {\far} on skewed datasets using \textbf{TPC-H Skew}
        benchmark~\cite{tpch_skew_linux}. We generate four datasets with SF=30 and different zipf factors ranging from 0.5 to 3.

      \begin{table*}
    \begin{minipage}[b]{0.66\linewidth}
    \centerline{\includegraphics[width=\linewidth]{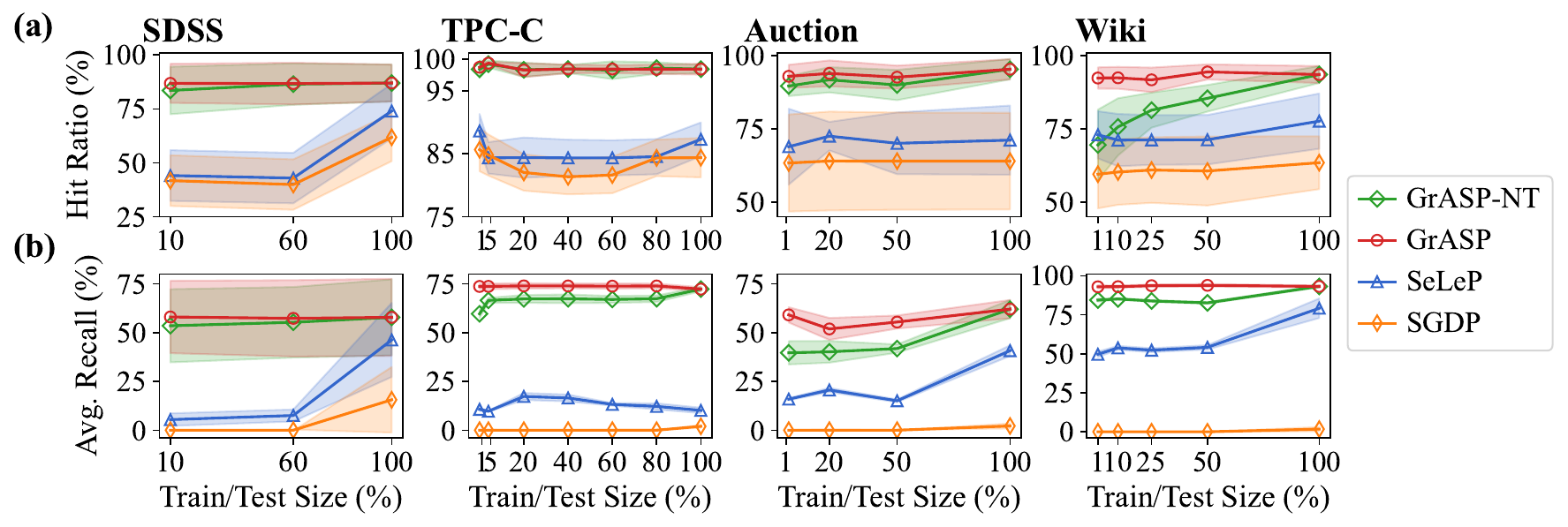}}
    \vspace{-1.2em}
    \captionof{figure}{(a) Hit ratio and (b) average recall with 95\% confidence interval in generalization tests. X-axis presents the size ratio of the train and test datasets.}
    \label{fig:generalization1}
    \end{minipage}\hfill
    \begin{minipage}[b]{0.32\linewidth}
    \centering
        \caption{Average Miss Coverage of {\far} and Baselines on Generalize Datasets and Two Selected Size Ratio (SR).}
        \vspace{-1em}
        \scalebox{0.87}{
        \begin{tabular}{>{\centering\arraybackslash}p{.84cm}>{\centering\arraybackslash}p{.35cm}>{\centering\arraybackslash}p{.65cm}>{\centering\arraybackslash}p{1.25cm}>{\centering\arraybackslash}p{.7cm}>{\centering\arraybackslash}p{.6cm}}
        \small{Dataset}       & SR & {\far} & \small{{\far}-NT} & {SeLeP} & {SGDP}\\ \hline
        \multirow{2}{*}{\small{SDSS}}& 10\%         & \textbf{79.41}       & \underline{75.42}        & 9.59  & 7.23  \\
        & 60\%         & \textbf{78.93}      & \underline{78.66}        & 7.6  & 4.6  \\ \hline
        \multirow{2}{*}{\small{TPC-C}}& 20\%         & \underline{86.35}       & \textbf{87.1 }       & 1.33  & 1.03  \\
        & 60\%         & \textbf{87.62}       & \underline{87.6 }       & 1.16  & 1.56  \\ \hline
        \multirow{2}{*}{\small{Auction}}& 20\%         & \textbf{77.97}       & \underline{69.77 }       & 36.88  & 2.75  \\
        & 50\%         & \textbf{71.42}       & \underline{63.59}        & 18.88  & 2.7  \\ \hline
        \multirow{2}{*}{\small{Wiki}}& 10\%         & \textbf{84.18 }      & \underline{55.42}        & 40.31  & 3.74  \\
        & 50\%         & \textbf{88.27}       & \underline{70.68}        & 33.5  & 2.1  \\ \hline
        \end{tabular}

    }
        \label{tab:general_coverage}
    \end{minipage}
    \vspace{-1.3em}
\end{table*}

    \vspace{-0.4em}
    \subsection{Baselines}\label{sec:baselines}
        {\far} is compared against traditional prefetchers employed in mainstream DBMSs, and SOTA learning-based data prefetchers. 
    
    \begin{itemize}       
        \item \textbf{Lookahead (LA)} \cite{smith1978lookahead}: A simple prefetcher, used in many DBMSs, sequentially fetches blocks after the accessed ones.
        
        \item \textbf{Random Readahead (RandR)} \cite{opdenacker2007readahead}: If a predefined number ($l_{RR}$) of an extent blocks are accessed within its LBA trace window, the prefetcher fetches the entire extent.
        
        \item \textbf{Naïve prefetcher}~\cite{naive}: Fetches blocks by repeatedly adding the most frequent delta to the last accessed LBA.
        
        \item \textbf{SGDP} \cite{2023sgdp}: This SOTA prefetcher models interactions among delta streams with a weighted directed graph and learns delta patterns using a gated graph neural network (GGNN).

        \item \textbf{SeLeP}\cite{selep}: This SOTA database prefetcher partitions and fetches blocks based on the interdependencies of their data.
    \end{itemize}

   \new{Since {\far}'s prefetch size changes dynamically, we bound it to $k$ blocks for fair comparison. LA, Naïve, and SGDP are extended to prefetch $k$ blocks instead of one. The RandR model, originally implemented in MySQL Server, uses default settings. SeLeP fetches $k/\text{pSize}$ partitions, where \textit{pSize} is the partition size, and \textit{pSize} = 128 is reported to optimally balance cache utilization and performance.}

\vspace{-0.4em}    
    \subsection{Metrics}\label{sec:metrics}

        \new{Our experiments evaluate multiple performance metrics. Hit ratio (Equation \hyperref[tab:metric_formulas]{6-1}) measures the proportion of cache hits to total accesses, reflecting overall cache effectiveness. Prefetch recall (Equation \hyperref[tab:metric_formulas]{6-2}) assesses the accuracy of immediate predictions, i.e., how many blocks are prefetched for the next step. Since a system without prefetching (NP) achieves some hits through block reaccesses, miss coverage (Equation \hyperref[tab:metric_formulas]{6-3}) measures the fraction of NP cache misses eliminated by the prefetcher, isolating its impact.}
        \vspace{0.05em}

        \begin{center}
            \scalebox{0.88}{
            \begin{tabular}{lll}
                Hit Ratio = $\displaystyle\frac{\text{Hits}}{\text{Hits} + \text{Misses}}$  (6-1) 
                &
                Recall = $\displaystyle\frac{\text{Correct Prefetches}}{\text{Accessed Blocks}}$  (6-2) &
                \\[1.5em] 
                
                & \hspace{-8em} \shortstack{Miss\\ Coverage} = $\displaystyle\frac{\text{Misses}_{NP} - \text{Misses}}{\text{Misses}_{NP}}$ & \hspace{-3.27em} (6-3)
                \\[1em] 
            \end{tabular}\label{tab:metric_formulas}
            }
        \end{center}

        \vspace{-0.4em}
      \new{We evaluate I/O improvements using the relative I/O time defined in~\cite{selep} (Equation~\hyperref[tab:relative_exec_time]{7}), where $t\_io_{pr}$ is the I/O time of the system with $pr$ prefetcher, and $t\_io_{NP}$ is the NP I/O time. Relative I/O reflects the reduction in storage access. We also report throughput and 95th percentile query latency to assess  the overall performance impact.}

        \begin{center}
            \scalebox{0.95}{
            \vspace{-1.5em}
            \begin{tabular}{>{\raggedleft\arraybackslash}p{5.75cm}>{\raggedleft\arraybackslash}p{1.9cm}} 
                        $\text{Relative}\ t\_io_{pr} = \displaystyle\frac{t\_io_{pr}}{t\_io_{NP}}$ & (7) \\
            \end{tabular}\label{tab:relative_exec_time}
            }
        \end{center}

\vspace{-0.3em}
\section{Experimental Results}
    We evaluate {\far} using various training and testing workloads, and examine its performance through the following key questions:
    \vspace{-0.3em}
    \begin{itemize}
        \item How does {\far} generalize its prediction on an enlarged dataset compared to other learning-based baselines?(\S \ref{sec:generalization_result})
        \item How does {\far} improve database performance compared to traditional and SOTA baselines?(\S \ref{sec:analytical_transactional_results})
        \item How does {\far} perform on skewed datasets, {modified physical schema, and shifting workloads?}(\S \ref{sec:skewed})
        \item What is the time complexity of {\far}?(\S \ref{sec:time_analysis})
        \item How do hyperparameters affect {{\far}\small{'s}} performance?(\S \ref{sec:sensitivity})
    \end{itemize}
    
    \vspace{-0.7em}

    \begin{figure*}
        \centerline{\includegraphics[width=0.98\textwidth]{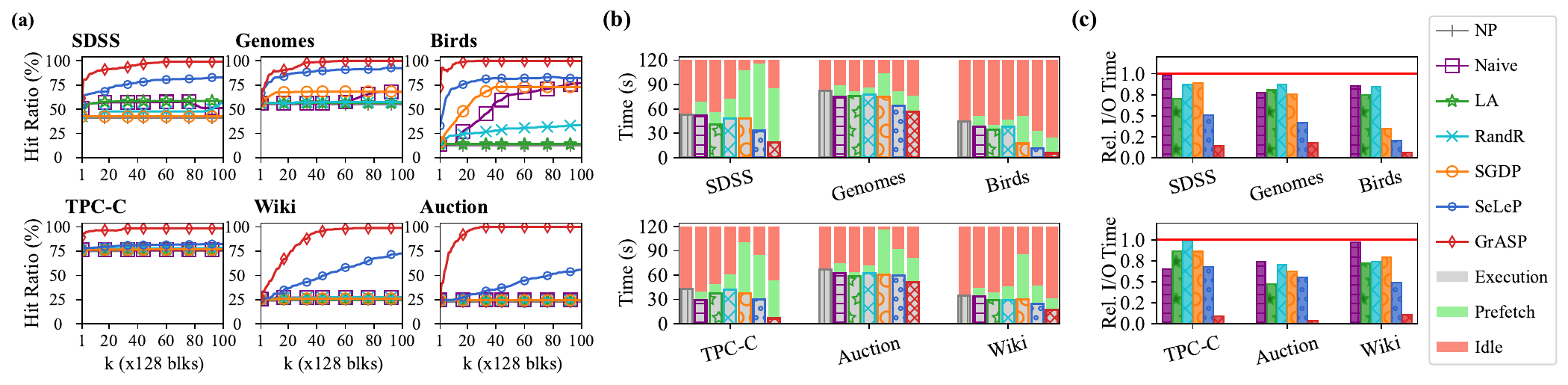}}
        \vspace{-1.5em}
        \caption{{(a) Hit ratio across prefetch sizes ($k$); (b) breakdown of execution, prefetch, and idle time in simulations at 5~qps with $k=50$; and (c) relative I/O time for the same simulations. Analytical workloads are on top and hybrid ones are below.}}
        \label{fig:analytical_transactional_hit_time}
        \vspace{-1.25em}
    \end{figure*}

    \subsection{Generalization Experiment}\label{sec:generalization_result}
 \new{This section evaluates the generalization of learning-based prefetchers on four datasets by training them on smaller databases and testing on larger ones (\S\ref{sec:generalization_workload}). After initial training, {\far} and SeLeP are fine-tuned using 5000 queries from the target database; {\far}-NT (no tuning) is included to isolate the impact of tuning.}

  \new{Figure~\ref{fig:generalization1} shows (a) hit ratio and (b) average recall across varying train-to-test dataset size ratios, using $k=50$ (\S\ref{sec:correctness}). A 100\% size ratio indicates training and testing on the same database, where prefetchers achieve their best performance. The results are averaged over 8 workload sessions with a 95\% confidence interval (CI).}

   Figure~\ref{fig:generalization1}(a) shows that {\far} consistently achieves the highest hit ratio across datasets, \new{reaching over 92\% even with minimal training on hybrid workloads. Its confidence intervals remain below 10\% in all cases except SDSS, where greater variability in query templates and access patterns leads to wider variation.}

    While collecting block access requests from the training datasets, we observed that the DBMS generates different query plans for similar queries across different SFs, even with identical indexes and schema. For instance, in the Auction dataset, the access patterns in SF 10 and 50 are similar, while SF 25 and 50 differ. Despite these variations, {\far} maintains robust performance across scales.
    
    {\far}-NT achieves hit ratios similar to {\far} in most tests, as frequent deltas in smaller training datasets often overlap with those in the target dataset, allowing effective prefetching. In addition, the query representations and block encodings remain similar across training and test datasets due to comparable data distributions and access patterns, enabling the model to generalize effectively without fine-tuning. \new{This results in stable performance and consistent CIs for {\far}, largely unaffected by dataset size ratio. In contrast, {\far}-NT exhibits greater variability and wider CIs, especially in the Wiki dataset, where diverging delta distributions and high block access rates make prediction without tuning more challenging.}

    Despite its adaptability, SeLeP underperforms relative to {\far}, especially in hybrid workloads. Even at a 100\% size ratio, some test blocks are absent from the training set, limiting SeLeP’s ability to generalize and achieve high accuracy. This limitation persists across all ratios, contributing to consistently low performance and a similar CI. The 5000 tuning queries are also insufficient to capture complex access patterns, especially in SDSS. In contrast, {\far} effectively uses these queries to refine delta values and adjust predictions.

    SGDP, utilizing the delta modeling method, is designed to handle size increases. However, it consistently fails to generalize and struggles with accuracy even at a 100\% size ratio due to its reluctance to prefetch under uncertainty. This highlights the limitation of relying solely on LBA-based information for access prediction in complex workloads. Additionally, we observed its recursive delta prediction approach is inefficient and significantly increases prediction time.

\begin{table}[h!]
    \caption{Average Recall (Rec) and Miss Coverage (MC) in Analytical and Hybrid Workloads With $k=50$.}
    \vspace{-1.55em}
    \begin{center}
    \scalebox{0.80}
    { 
    \begin{tabular}{l>{\centering\arraybackslash}p{.42cm}>{\centering\arraybackslash}p{.53cm}>{\centering\arraybackslash}p{.33cm}>{\centering\arraybackslash}p{.35cm}>{\centering\arraybackslash}p{.33cm}>{\centering\arraybackslash}p{.51cm}>{\centering\arraybackslash}p{.36cm}>{\centering\arraybackslash}p{.43cm}>{\centering\arraybackslash}p{.36cm}>{\centering\arraybackslash}p{.42cm}>{\centering\arraybackslash}p{.33cm}>{\centering\arraybackslash}p{.4cm}}
         & \multicolumn{2}{c}{SDSS} & \multicolumn{2}{c}{Genomes} & \multicolumn{2}{c}{Birds} & \multicolumn{2}{c}{TPC-C} & \multicolumn{2}{c}{Auction} & \multicolumn{2}{c}{Wiki} \\ \cline{2-13}
        \small{Method}              & Rec   & MC    & Rec   & MC    & Rec   & MC    & Rec   & MC    & Rec   & MC     & Rec   & MC  \\ \hline
        \small{Naive}               & 2.4   & 15.3  & 1.2   & 26.4  & 1.12  & 41.7  & 2.6   & 8.1   & 0     & 0.2    & 0  & 1.02  \\ 
        \small{LA}                  & 9.6   & 27.9  & 1.3   & 3.3   & 1.9   & 12.1  & 0.8   & 1.02  & 2.2   & 3.9    & 7.9   & 2.24  \\ 
        \small{RandR}               & 3.8   & 9.3   & 1.18  & 2.06  & 27.6  & 15.3  & 0     & 0.4   & 0     & 0.6    & 8.23  & 2.3   \\ 
        \small{SGDP}                & 1.5   & 1.97  & 1.7   & 27.5  & 4.1   & 66    & 1.4   & 1.97  & 1.6  & 1.3    & 0.9  & 0.6  \\ 
        \small{SeLeP}               & \underline{38.25} & \underline{60.3}  & \underline{82.8}  & \underline{76.65}  & \underline{16.8}  & \underline{78.7}  & \underline{7.7} & \underline{17} & \underline{42.1}  & \underline{11.4}    & \underline{77.9}  & \underline{17.2} \\ 
        {\far}                & \textbf{90.8}  & \textbf{91.5} & \textbf{83.6}  & \textbf{89.7}  & \textbf{99.1}  & \textbf{99.8} & \textbf{72.2}  & \textbf{87.4} & \textbf{61.9}    & \textbf{95.3} & \textbf{91.9}  & \textbf{91.2} \\ \hline
    \end{tabular}
    }
    \end{center}
    \label{tab:analytical_transactional_recall_coverage}
    \vspace{-1.75em}
\end{table}

   \new{Figure \ref{fig:generalization1}(b) demonstrates that {\far} achieves the highest average recall with a narrow confidence interval. At a prefetch size of $k=50$, this metric reflects the average per-query hit ratio in a 400MB cache. Therefore, higher recall is critical when the memory budget for prefetching is limited. Achieving a high hit ratio with low recall means that cache hits are primarily due to blocks prefetched in previous steps that were not promptly accessed.}

    Table \ref{tab:general_coverage} shows miss coverage of prefetchers in two different size ratios. {\far} consistently outperforms the baselines, achieving an average miss coverage of 83.75\% with fine-tuning and 71.25\% without it. In contrast, SeLeP and SGDP fail to surpass 40\% miss coverage. A low miss coverage indicates an inability to anticipate and prefetch upcoming accesses not already present in the cache.

    \vspace{-0.5em}
    \subsection{Analytical and Transactional Experiment}\label{sec:analytical_transactional_results}
    {This section compares {\far} with baselines (\S\ref{sec:baselines}) on analytical and hybrid workloads (\S\ref{sec:datasets}) over various performance metrics (\S\ref{sec:metrics}).}

    \vspace{-0.25em}
    \subsubsection{Correctness and Coverage}\label{sec:correctness}
        {Figure~\ref{fig:analytical_transactional_hit_time}(a) shows hit ratios across various prefetching $k$, guiding our choice of $k$ for other experiments. Table~\ref{tab:analytical_transactional_recall_coverage} reports average recall and miss coverage at $k=50$.}

    Analytical workloads, which do not modify data, are generally easier to predict, especially in smaller datasets like Birds, where LBA-based prefetchers perform well. However, as dataset size or block access rate increases, LBA-based and traditional prefetchers struggle to fill the cache effectively. In contrast, semantic prefetchers excel, with {\far} consistently outperforming all baselines.  
    
   \new{Hybrid workloads challenge prefetchers, especially LBA-based and traditional ones. While SeLeP struggles with dominant transactional queries, {\far} performs well across all workload types. It achieves its best hit ratio near $k=20$ for low-access workloads (Birds, TPC-C) and $k=40$ for high-access workloads (Wiki, Auction, Genomes). Since other baselines perform similarly around $k=50$, we use this value in all evaluations.}

    \autoref{tab:analytical_transactional_recall_coverage} highlights the superiority of semantic prefetchers in terms of prefetching recall and miss coverage, where {\far} is always the best and SeLeP ranks second. However, {\far} demonstrates significantly better performance than SeLeP in most datasets, with differences of up to 83\% in recall and 84\% in miss coverage.

    \begin{figure*}
        \centerline{\includegraphics[width=0.925\linewidth]{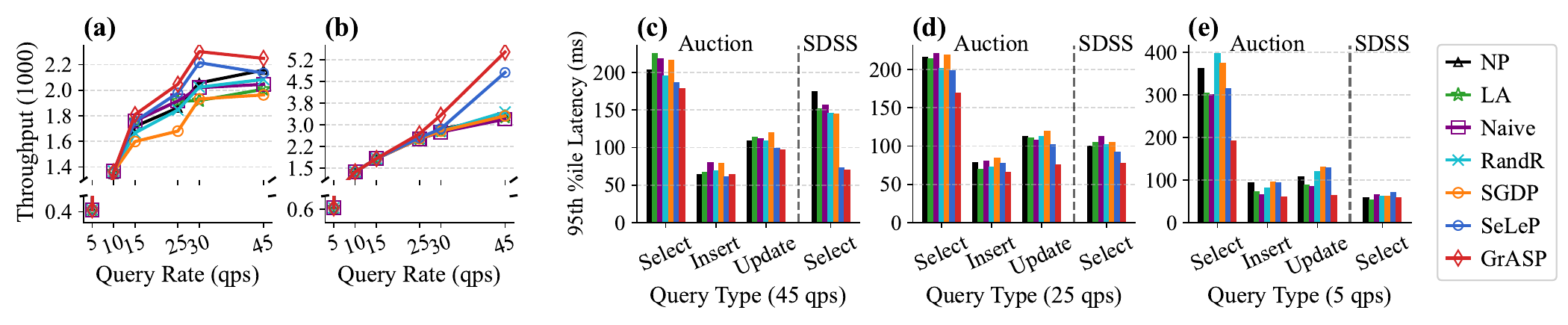}}
        \vspace{-1.3em}
        \caption{{Throughput on (a) Auction and (b) SDSS; 95th percentile latency at (c) 45 qps (25ms max delay), (d) 25 qps (50ms max delay), and (e) 5 qps (250ms max delay). Higher throughput and lower latency are better.}}
        \label{fig:throughput}
        \vspace{-1.3em}
    \end{figure*}

    \vspace{-0.4em}
    \subsubsection{\new{Runtime Impact}}
    \new{Prefetching reduces query response time by reducing I/O delays, as computation time is unaffected. To evaluate prefetchers' effectiveness, we report the execution time breakdown and corresponding relative I/O time in Figure~\ref{fig:analytical_transactional_hit_time}(b) and \ref{fig:analytical_transactional_hit_time}(c). 
    }

    \new{Figure~\ref{fig:analytical_transactional_hit_time}(b) breaks down simulation time at 5 qps (queries per second, or query rate), corresponding to a maximum query interarrival delay of 250 ms. It shows total execution time (patterned gray), total prefetch time (green), and system idle time (pink). The simulations run until the full workload is processed, while prefetching is \emph{non-blocking} and stops as soon as the next query arrives.}

    \new{{\far} consistently achieves the greatest execution time reduction across all datasets. Compared to NP, it saves over 85\% on Birds and TPC-C, 65\% on SDSS, 50\% on Wiki, and up to 30\% on others. It also outperforms SeLeP by 9–55\%, with the largest gains on TPC-C (55\%), SDSS (28\%), Wiki (22\%), and Birds (14\%).}

    \new{{\far}'s prefetch time is lower than other learning-based methods, as it dynamically estimates and adjusts prefetch size (bounded by~$k$) rather than using a fixed size. In our tests, {\far} prefetched the same or fewer blocks per query than SeLeP—averaging 19.13\% fewer overall and up to 93.42\% fewer in some queries. With lower prefetch overhead, {\far} reduces prefetch time by 37\% on Wiki, 19\% on SDSS, and up to 14\% on other datasets.}

\new{Note that a non-blocking prefetcher utilizes idle time \emph{without} adding overhead, so the true end-to-end latency excludes prefetch time. As shown in Figure~\ref{fig:analytical_transactional_hit_time}(b), {\far} effectively uses idle periods to prefetch relevant blocks, resulting in lower end-to-end latency.}

   \new{\autoref{fig:analytical_transactional_hit_time}(c) reports the relative I/O time corresponding to the simulations in \ref{fig:analytical_transactional_hit_time}(b), where a value of 1 indicates the maximum I/O time (NP), and 0 represents the ideal case with all data served from the cache. For reference, a value of 0.2 shows an 80\% I/O reduction.}

    Across all tests, {\far} achieves the lowest relative I/O time, reducing I/O delays by up to {96\%} in analytical workloads and up to {94\%} in hybrid ones. SeLeP ranks second but struggles with hybrid workloads, achieving less than a 51\% I/O reduction even in the ones with mainly analytical queries. Traditional prefetchers perform close to NP, offering at most a 48\% improvement in I/O time.
    
   \new{Although higher miss coverage generally reduces I/O time, the actual performance depends on the physical location of the accessed blocks. Some blocks may need more movement or processing, leading to variation in I/O times even for systems with similar statistics.}

    \begin{figure}
        \centering
        \includegraphics[width=0.96\linewidth]{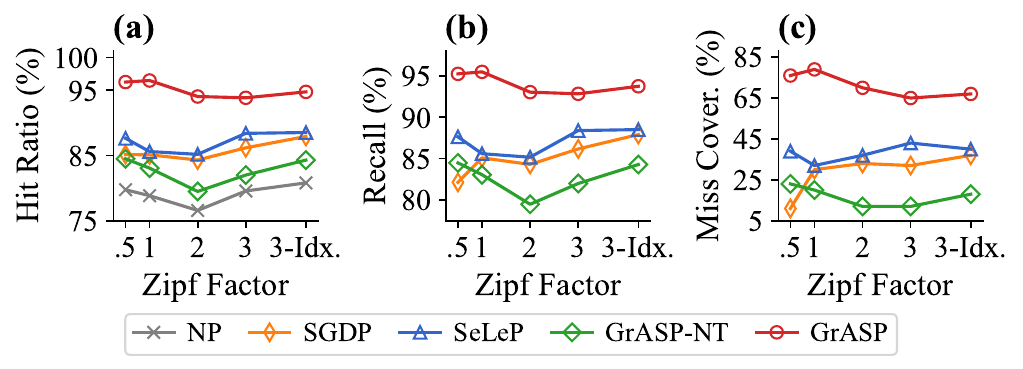}
            \vspace{-1.4em}            
            \captionof{figure}{(a) Hit ratio, (b) recall, and (c) miss coverage in original and indexed TPC-H Skew with various $z$ ($k$=1500).}
            \label{fig:skewed_hit}
            \vspace{-1.5em}   
    \end{figure}

    \vspace{0.5em}  
    \textit{Throughput and latency.} \new{We extend our runtime analysis by evaluating the prefetcher impact on system throughput and 95th percentile query latency across varying arrival rates. Figure~\ref{fig:throughput} shows results from 120s simulations on the Auction hybrid high-access workload (Table~\ref{tab:db_workload}) and the SDSS analytical workload. The tested rates correspond to maximum interarrival delays ($d$) ranging from 25 to 250~ms, with actual delays sampled from $[d/2, d]$ and biased toward $d$, averaging $7d/8$—similar to the skewed arrival in \cite{dist}.}

   \new{Figure~\ref{fig:throughput}(a) shows throughput on the Auction dataset. At low query rates, throughput remains similar across methods, as simple or repeated queries complete quickly, balancing more complex or longer-running ones. Nonetheless, \autoref{fig:analytical_transactional_hit_time}(b) reveals notable differences in execution times across prefetchers for the same number of queries, with {\far} completing them up to 85\% faster. As load increases, some prefetchers degrade throughput by fetching irrelevant blocks, falling below NP, while {\far} maintains the highest throughput. At 45~qps, the system saturates and throughput drops due to minimal available time ($<$25~ms) for prefetching.}

    \new{The throughput results of the SDSS dataset in Figure~\ref{fig:throughput}(b) follow a similar trend. However, due to the lower block access rate in SDSS, {\far} outperforms others even at 45 qps, executing 20–85\% more queries. At higher qps however, all prefetchers fail to keep up.}

    \new{Figures~\ref{fig:throughput}(c-e) show 95th percentile latency per query type\footnote{Auction contains very few DELETE queries—only 8 among 100k training queries.} at 45, 25, and 5 qps, matching $d$ values of 25, 50, and 250 ms. {\far} consistently delivers lower latency with up to 57\% for selection and 32\% for transactional queries, even under high loads. In contrast, SeLeP shows less stable latency, and some prefetchers degrade performance by polluting the cache. Note that at higher qps, prefetchers execute a different number of queries, which can affect their latency.}

    \vspace{-0.4em}
    \subsection{Adaptivity Experiment}\label{sec:adaptivity}
        {Real-world database workloads often exhibit non-uniform data dist- ributions and changing query patterns, as user interests shift over time~\cite{ResearcherGuide, skew1}. In addition, variations in available resources or physical schemas may cause the database to select different query plans for identical queries at different times. In this section, we evaluate how {\far} performs under these realistic and dynamic conditions.}

    \begin{figure}
        \centerline{\includegraphics[width=0.96\linewidth]{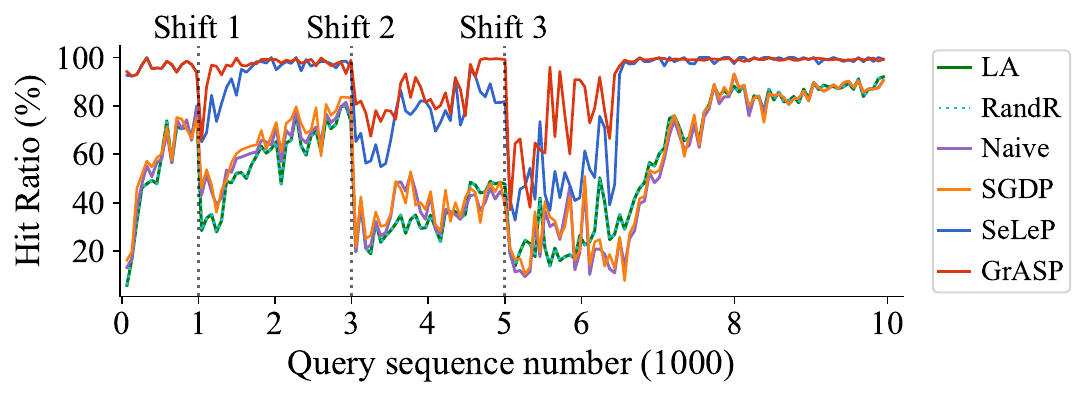}}
        \vspace{-1.45em}
        \caption{Hit ratio for consecutive non-overlapping 60-query batches in a shifting SDSS workload with $k$\small{=50}.}
        \label{fig:adaptivity}
        \vspace{-1.55em}   
    \end{figure}

    \begin{figure*}
        \includegraphics[width=\linewidth]{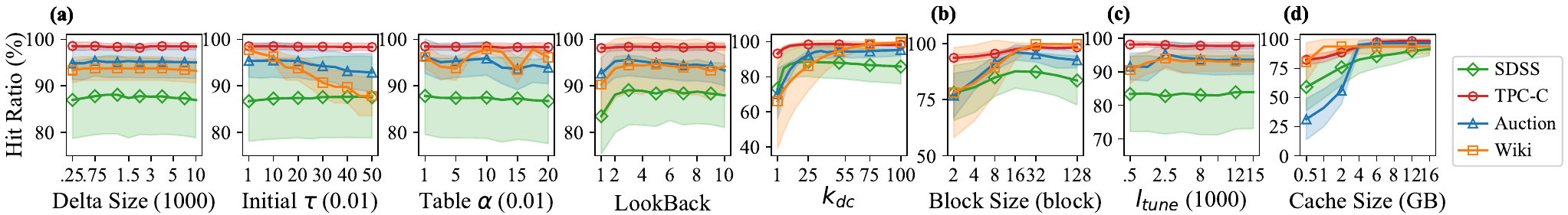}
        \vspace{-1.9em}
        \caption{Effect of (a) model config, (b) block size, (c) tuning query count, and (d) cache size on {\far}'s hit ratio with 95\% CI.}
        \label{fig:sensitivity} 
        \vspace{-1.25em}
    \end{figure*}

    \vspace{-0.4em}
    \subsubsection{\textbf{Skewed Dataset}}\label{sec:skewed}
    \new{This subsection evaluates the prefetchers on skewed TPC-H datasets with varying Zipf factors ($z$). Models are trained on a dataset with SF=10 and z=0.5, and tested on datasets with SF=30 and z=\{0.5, 1, 2, 3\}.  In TPC-H Skew, higher $z$ values correspond to more pronounced skewness, where some customers place more orders and certain parts are ordered more frequently. The test workloads run similar queries across datasets, with {\far} and SeLeP fine-tuned using 500 queries from the target dataset. }
    
   \new{Figure~\ref{fig:skewed_hit} presents (a) hit ratio, (b) recall, and (c) miss coverage of the skew tests. Since TPC-H Skew queries access significantly more blocks than other workloads, achieving high performance demands prefetching a much larger number of blocks ($k=1500$).}

    \new{Figure \ref{fig:skewed_hit} shows that {\far}, after tuning on a few queries, achieves over 93\% hit ratio and recall, and up to 80\% miss coverage. Data skewness does not drastically impact {\far} as it includes LBA details in its input context, while SeLeP, relying solely on semantics, is more affected and performs similarly to the LBA-based SGDP. Since both semantic and LBA contexts shift in these tests, {\far}-NT performs significantly worse and fine-tuning is critical.}

    \vspace{-0.2em}
    \subsubsection{\textbf{{Modified Physical Schema}}}\label{sec:index}
        {We investigate the impact of physical design by evaluating the prefetchers on the TPC-H Skew dataset (with z=3), augmented with new indexes derived from HMAB~\cite{hmab}, a SOTA database tuning tool. The results of this test are labeled "3-Index" in \autoref{fig:skewed_hit}. By reducing the total number of blocks accessed, indexes increase block reaccess rates and improve prediction accuracy, resulting in higher hit ratios across all systems, including NP. While outperforming all baselines, {\far} maintains consistent performance across physical design changes and query plan variations due to its plan-agnostic query encoding.}
 \begin{table}[htbp]
        \vspace{0.2em}
        \caption{Average time of encoding {\small{100}} blocks, model initializing, delta prediction, fetching blocks, tuning the generalized model ($l_{tune}$={\small{5000}}), and training of the semantic prefetchers. }
        \vspace{-0.5em}
        \begin{center}
            \scalebox{0.81}{
            \begin{tabular}{|c|c|c|c|c|c|}
                \hline
                 Operation                                          & SDSS      & Wiki      & Auction   & TPC-C     & \small{TPC-H Skew} \\ \hline
                \centered{Encoding of 100\\blocks (s) one-off}      & 20.71     & 3.6       & 59.17     & 49.04     & 75.89 \\ \hline
                \centered{Model initialization \\ (s) one-off}     & 1132.11   & 55.94     & 55.19     & 395.49    & 160.6 \\\hline
                \centered{Delta prediction \\ per query (ms)}       & 3.06      & 6.57      & 4.57      & 2.4       & 27.29 \\ \hline
                \centered{Block prefetching \\ per query (s)}       & 0.25      & 0.06      & 0.36      & 0.15      & 2.2 \\ \hline
                \centered{Fine-tuning (s)}                          & 23.25     & 27.14    & 21.6      & 21.8      & 40.21 \\ \hline
                \centered{\textcolor{black}{Training (s)}}           &\textcolor{black}{570.17}     &\textcolor{black}{ 27.97}    &\textcolor{black}{ 37.41}      & \textcolor{black}{250.86}      & \textcolor{black}{27.78} \\ \hline
            \end{tabular}
         }
         \label{tab:times}
        \end{center}
    \vspace{-1.5em}
    \end{table}

    \vspace{-0.2em}
    \subsubsection{\textbf{{Shifting Workload}}}\label{sec:shift}
        {To evaluate adaptivity, we simulate evolving workloads on the SDSS dataset. The cache is first warmed up, followed by three staged shifts: at sequence number (SN)=1000, 25\% of blocks are unseen; at SN=3000, 40\% of blocks, 50\% of tables, and 50\% of templates are new; and at SN=5000, all tables change and unseen blocks are accessed using entirely new templates. {\far} updates delta classes every $l_{tune} = 500$ queries, fine-tuning only if deltas change, with 5.54s average overhead. SeLeP also tunes every $l_{tune}$ queries, taking 46.42s due to the costly repartitioning.}

        {Figure~\ref{fig:adaptivity} shows hit ratios of batches of 60 consecutive queries up to SN=10000. {\far}'s hit ratio drops less sharply and recovers faster, maintaining a relatively consistent performance—especially in the first shift, where it improves before tuning. After the final shift, the fully unseen workload increases model uncertainty, leading {\far} to skip prefetching for some queries, while SeLeP issues inefficient random prefetches that occasionally succeed by chance. Upon tuning at SN=5500, {\far}'s average hit rate increases from 53\% to 81\%, while SeLeP peaks below 75\% despite fine-tuning. Due to the stochastic nature of the workload, all prefetchers struggle to stabilize until SN=7000, where only {\far} and SeLeP converge.}

    \vspace{-0.25em}
    \subsection{Time Analysis}\label{sec:time_analysis}

    \new{Table \ref{tab:times} reports the {\far}'s time overhead across datasets. The block encoding step of model initialization is reported separately as it heavily depends on the train workload size. It is also influenced by the number of columns and their data types; datasets with fewer columns (Wiki) or primarily numerical data (SDSS) have lower overhead. Model initialization time scales with the number and complexity of training queries, with SDSS and TPC-H Skew showing higher overheads due to more complex query statements.}

    \new{Delta prediction involves context creation and model inference, both influenced by query complexity and block selectivity. Thus, TPC-H Skew, which accesses 30k blocks with complex queries, exhibits the highest prediction time. However, this time stays within the millisecond range and does not impact database interactivity.  }

    Block prefetching calculates block LBAs using predicted deltas and retrieves them from storage. This time decreases if the block is already in the cache and increases with a higher number of fetches. Except for TPC-H Skew, which has a high block access rate, prefetching is completed in under 400ms, ensuring interactivity.

    \new{Fine-tuning time includes delta adjustment and prediction model tuning. TPC-H Skew, with its higher query encoding time, has the longest tuning time. However, tuning takes under a minute and can be run asynchronously alongside the main prefetcher functions.}
      
    \new{Unlike SeLeP, {\far}'s fine-tuning does not require encoding new blocks. SeLeP must re-encode all newly inserted blocks and assign them to partitions during each tuning event, resulting in overheads orders of magnitude higher than those of {\far}.}

    \new{Table~\ref{tab:times} shows {\far}'s training times. Compared to {\far}, SeLeP’s larger model with two additional fully connected layers, and SGDP’s GGNN requiring costly graph message-passing, increase training costs by up to 9.7 and 418.4 times, respectively. For comparison, SeLeP's training times are 5573, 60, 30, 1369, and 31 s, while SGDP's times are 4340, 147, 1245, 449, and 11715 s, respectively.}

    \vspace{-0.75em}
    \subsection{Sensitivity Analysis}\label{sec:sensitivity}
    \new{To assess the impact of parameters, we evaluate {\far} under different settings on the SDSS and hybrid workloads, reporting hit ratios with 95\% CI. The results confirm that the settings described in \S5.1 deliver stable and robust performance across workloads.}

    \new{\textit{Model Parameters} impact is shown in Figure \ref{fig:sensitivity}(a). Small \textit{delta size ($ds$)} fail to cover all frequent deltas, reducing performance, while large $ds$ introduce too many classes, making accurate predictions harder. A high \textit{threshold} $\tau$ applies overly strict table selection, and extreme \textit{table $\alpha$} values distort $\tau$, both lowering performance.}

    \new{Using query history improves predictions, as shown by higher hit ratios for models with $n_{lookback}>1$. However, excessively large histories increase model complexity and reduce performance. For datasets with high block access rates, a larger $k_{dc}$ boosts performance, but overly high values risk selecting irrelevant blocks.}

    \textit{Block size.} \new{Figure \ref{fig:sensitivity}(b) illustrates that increasing {block size} reduces the number of deltas, improving prediction accuracy. However, very large block sizes populate the cache with unused data.}

    \new{\textit{Tuning query count ($l_{tune}$).} Figure~\ref{fig:sensitivity}(c) shows {\far} adapts deltas and predictions after tuning on at least 2,500 queries, or as few as 500 queries for datasets with low block access like TPC-C.}

    \new{\textit{Cache Size} impact is shown in Figure~\ref{fig:sensitivity}(d). Larger caches preserve longer access histories, increasing block reaccess chances and extending the impact of prefetches. {\far} maintains a high hit ratio with a 2GB cache across datasets from 16GB to 155GB.}

\vspace{-0.2em}    
\section{Conclusion}
    This paper presents {\far}, a learning-based semantic prefetcher designed to enhance database interactivity by leveraging both LBA patterns and data semantics. {\far} combines  queries LBA-Delta with their encoded semantics to predict future delta values and optimize prefetching hit ratio across a diverse range of workloads. Our evaluation on analytical and transactional workloads demonstrates that {\far} significantly improves performance, outperforming SOTA methods with up to {45}\% higher hit ratio, \new{{60}\% lower I/O time, and 55\% lower execution latency}. Additionally, our experiments on enlarged datasets demonstrate that {\far}, through delta modeling and lightweight fine-tuning, generalizes its high performance to datasets up to \new{250$\times$} larger, and with different skewed data distributions—capabilities not achievable by SOTA semantic prefetchers.

\balance
\bibliographystyle{ACM-Reference-Format}
\bibliography{bib}

\end{document}